\documentclass[conference]{IEEEtran}
\IEEEoverridecommandlockouts
\pdfoutput=1
\usepackage[noadjust]{cite}
\usepackage{amsmath}
\usepackage{breqn}
\usepackage{array}
\usepackage{mdwmath}
\usepackage{amssymb}
\usepackage{algorithm}
\usepackage[noend]{algpseudocode}
\usepackage{eqparbox}
\usepackage{stfloats}
\usepackage{url}
\usepackage{tikz}
\usepackage{pgfplots}
\graphicspath{{../pdf/}{../jpeg/}}
\DeclareGraphicsExtensions{.pdf,.jpeg,.png}
\usepackage{epstopdf}
\usepackage{amsfonts}
\usepackage{tikz}

\newtheorem{lemma}{Lemma}
\newtheorem{theorem}{Theorem}
\newtheorem{remark}{Remark}
\newtheorem{corollary}{Corollary}
\newtheorem{assumption}{Assumption}

\begin{document}

\title{\huge Intelligent Reflecting Surface Assisted MISO Downlink: Channel Estimation and Asymptotic Analysis}   

\author{\IEEEauthorblockN{Bayan Al-Nahhas, Qurrat-Ul-Ain Nadeem, and Anas  Chaaban}
\IEEEauthorblockA{School of Engineering, University of British Columbia, Kelowna, Canada. \\ 
Email:\{bayan.alnahhas, qurrat.nadeem, anas.chaaban\}@ubc.ca}
\thanks{This work is supported by the King Abdullah University of Science and Technology (KAUST) under Award No. OSR-2018-CRG7-3734.}}

\maketitle

\begin{abstract}

 This work makes the preliminary contribution of studying the asymptotic performance of a multi-user intelligent reflecting surface (IRS) assisted-multiple-input single-output (MISO) downlink system under imperfect CSI. We first extend the existing least squares (LS) ON/OFF channel estimation protocol to a multi-user system, where we derive minimum mean squared error (MMSE) estimates of all IRS-assisted channels over multiple sub-phases. We also consider a low-complexity direct estimation (DE) scheme, where the BS obtains the MMSE estimate of the overall channel in a single sub-phase. Under both protocols, the BS implements maximum ratio transmission (MRT) precoding while the IRS design is studied in the large system limit, where we derive deterministic equivalents of the signal-to-interference-plus-noise ratio (SINR) and the sum-rate.  The derived asymptotic expressions, which depend only on channel statistics, reveal that under Rayleigh fading IRS-to-users channels, the IRS phase-shift values do not play a significant role in improving the sum-rate but the IRS still provides an array gain.  Simulation results confirm the accuracy of the derived deterministic equivalents and show that under Rayleigh fading, the IRS gains are more significant in noise-limited scenarios. We also conclude that  the DE of the overall channel yields better performance when considering large systems.

\end{abstract}

\vspace{-.06in}
\section{Introduction}
\vspace{-.02in}

\label{Sec:Intro}

Fifth Generation (5G) technologies like massive multiple-input multiple-output (MIMO), small cells and millimeter wave communication not only consume a lot of energy but their performance is highly prone to losses in harsh propagation environments \cite{ML2}. To address these limitations, the concept of deploying intelligent reflecting surfaces (IRSs) in the environment has emerged in recent works, where the IRS is envisioned as a planar array of passive reflecting elements that can independently induce phase shifts onto the incident electromagnetic waves for performance enhancement  \cite{6G2, SRE1}.  


Several recent works jointly optimize the precoding at the base station (BS) and the phase-shifts applied by IRS elements, under different performance criteria of interest, for example: sum-rate maximization in \cite{Guo}, energy efficiency maximization subject to users' signal-to-interference-plus-noise ratio (SINR) constraints in \cite{huang}, transmit power minimization in \cite{Wu1, Wu_dis} and maximization of minimum SINR in \cite{annie}. The majority of these and other related works study the IRS performance under perfect channel state information (CSI) assumption.


CSI acquisition is a critical issue in IRS-assisted systems due to the passive nature of the IRS elements \cite{LS, LS1, CE_MU}. Moreover, the number of channels estimates needed to design IRS parameters is generally large, rendering the training overhead to be high. For a single-user system, the ON/OFF channel estimation scheme is proposed in \cite{LS} that serially develops least squares (LS) estimates of all IRS-assisted links. The protocol is improved in \cite{LS1} by allowing all IRS elements to be active during training. Both protocols require as many estimation sub-phases as the number of IRS elements. 

In this work, we consider a multi-user IRS-assisted downlink MISO system where like most other works, we assume the IRS-to-users channels to undergo independent Rayleigh fading \cite{huang, Wu_dis}. We then extend the ON/OFF estimation scheme from \cite{LS} to a multi-user system and derive the minimum mean square error (MMSE) estimates of all IRS-assisted channels over multiple sub-phases. Recognizing the large overhead imposed by this protocol, we also propose a direct estimation (DE) scheme in which the BS estimates the overall channel to each user in one sub-phase, instead of estimating the individual links over multiple sub-phases. This protocol, while not very practical in the non-asymptotic regime where we need estimates of all individual IRS-assisted channels to design the IRS \cite{LS}, can be a viable scheme for large systems, where the SINR and sum-rate performance approaches deterministic quantities that do not depend on instantaneous channels. 



The derived estimates are used to implement maximum ratio transmission (MRT) precoding at the BS. To design the IRS parameters, we resort to the asymptotic analysis of the sum-rate motivated by the large system sizes envisioned for future networks. Specifically, we develop deterministic equivalents of the SINR and sum-rate under both estimation protocols, which become tight in the large system limit \cite{Wag, ML2}. The deterministic equivalents under MMSE-ON/OFF protocol show some dependence on the values of IRS phase-shifts, which are optimized using projected gradient ascent requiring knowledge of only the channel statistics. The deterministic equivalents under DE  and perfect CSI do not show a notable dependence on IRS phase-shifts under Rayleigh fading. To explicitly study the asymptotic impact of IRS on the SINR, we apply the results to a special case from \cite{ML} and find that (i) IRS yields an array gain but no significant reflect beamforming gain under Rayleigh fading and (ii) the performance gain is high in noise-limited systems. Simulation results verify these insights and illustrate the excellent match yielded by deterministic equivalents. Moreover, we observe that if the IRS is designed using the sum-rate deterministic equivalent, the DE scheme becomes desirable due to the  reduction in training time. 


The rest of the paper is organized as follows. In Sec. \ref{Sec:Sys} the IRS-assisted system model and channel estimation protocols are outlined. Sec. \ref{Sec:Asym} presents the asymptotic analysis. Sec. \ref{Sec:Sim} provides simulation results and Sec. \ref{Sec:Con} concludes the paper.

 \vspace{-.04in}
\section{System Model} \label{Sec:Sys}
 \vspace{-.04in}

In this section, we present the transmission model and outline the MMSE-ON/OFF as well as the DE protocol. 

\begin{figure}[!t]
\centering
\includegraphics[width=.4\textwidth, height=.2\textwidth]{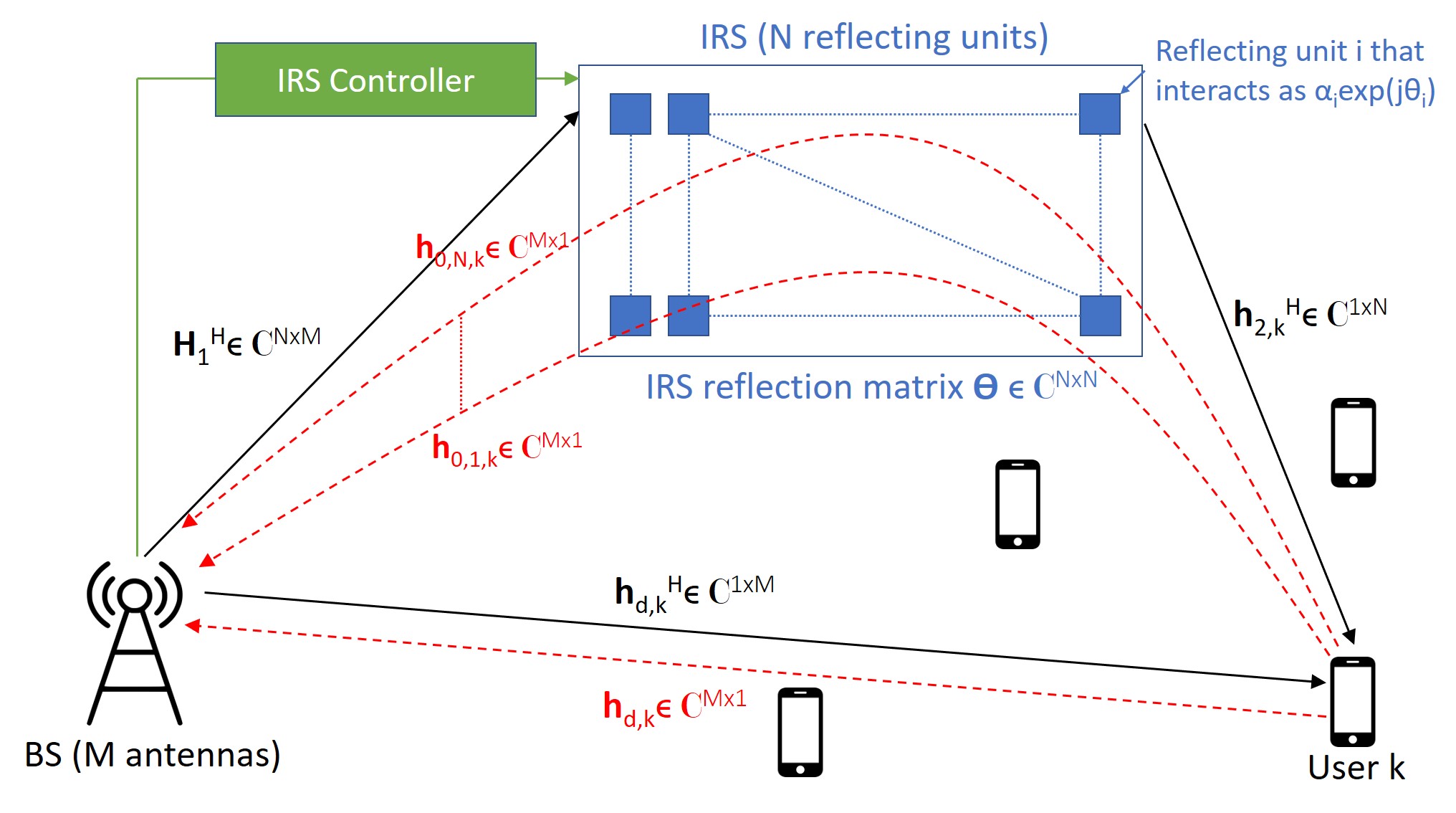}
\caption{Sketch of an IRS-assisted MISO system.}
\label{LIS_model}
\end{figure}

 \vspace{-.02in}

\subsection{Signal Model}

As shown in Fig. \ref{LIS_model}, a BS equipped with $M$ antennas communicates with $K$ single antenna users. This communication is assisted by an IRS composed of $N$ passive reflecting elements installed in the line-of-sight (LoS) of the BS. The IRS is equipped with a smart controller, which can communicate with the BS over a backhaul link for configuration. The received baseband signal $y_{k}$ at user $k$ is given as
\begin{align}
\label{model}
&y_{k}= \mathbf{h}_{k,\mathbf{v}}^{H} \mathbf{x}+n_k,
\end{align}
where the overall channel $\mathbf{h}_{k,\mathbf{v}}\in \mathbb{C}^{M\times 1}$ between BS and user $k$ is represented as
\begin{align}
\label{eq_ch1}
&\mathbf{h}_{k,\mathbf{v}}=\mathbf{h}_{d,k}+\mathbf{H}_{1}\boldsymbol{\Theta} \mathbf{h}_{2,k}
\end{align}
where $\mathbf{H}_{1}\in \mathbb{C}^{M\times N}$ is the LoS deterministic channel between the IRS and the BS represented as $\mathbf{H}_{1}=[\sqrt{\beta}_{1}\mathbf{h}_{1,1}, \dots, \sqrt{\beta}_{N}\mathbf{h}_{1,N}]$, where $\beta_1$ is the path loss factor and $\mathbf{h}_{1,n}$ is the channel between the BS and the $n^{th}$ IRS element. Also $\mathbf{h}_{2,k}\sim \mathcal{CN}(\textbf{0},\beta_{2,k} \mathbf{I}_N) \in \mathbb{C}^{N\times 1}$ and  $\mathbf{h}_{d,k}\sim \mathcal{CN}(\textbf{0},\beta_{d,k} \mathbf{I}_M)\in \mathbb{C}^{M\times 1}$ are the  Rayleigh fading channel vectors between user $k$ and the IRS, and  user $k$ and the BS respectively, where $\beta_{2,k}$ and $\beta_{d,k}$ are the channel attenuation coefficients. Finally $\boldsymbol{\Theta}=\text{diag}(\mathbf{v})\in \mathbb{C}^{N\times N}$ captures the response of the IRS, where $\mathbf{v}=[\alpha_1 \exp(j\theta_{1}), \alpha_2 \exp(j\theta_{2}),\dots, \alpha_N \exp(j\theta_{N})]^T\in \mathbb{C}^{N\times 1}$ is the IRS reflect beamforming vector, with $\theta_{n}\in[0,2\pi] $ being the induced phase-shift and $\alpha_{n} \in[0,1] $ being the given amplitude reflection coefficient of element $n$.  We denote $v_n=\alpha_n \exp(j\theta_n)$  in the remainder of this work. 

Moreover, $\mathbf{x}=\sum_{k=1}^{K} \sqrt{p_{k}} \mathbf{g}_{k} s_{k}$ is the transmit (Tx) signal vector, where $\mathbf{g}_{k} \in \mathbb{C}^{M\times 1}$, $p_{k}$ and $s_{k} \sim \mathcal{CN}(0,1)$ are the precoding vector, signal power  and data symbol for user $k$ respectively. The Tx vector satisfies the average constraint
\begin{align}
\label{p_cons}
& \mathbb{E}[||\mathbf{x}||^{2}]=\text{tr}(\mathbf{P} \mathbf{G}^{H} \mathbf{G}) \leq  P_{max},
\end{align}
where $P_{max}$ is the Tx power budget, $\mathbf{P}=\text{diag}(p_{1}, \dots, p_{K})$ and $\mathbf{G}=[\mathbf{g}_{1}, \mathbf{g}_{2}, \dots,\mathbf{g}_{K}]$. The BS utilizes the estimates of $\mathbf{h}_{k,\mathbf{v}}$, denoted as $\hat{\mathbf{h}}_{k,\mathbf{v}}$, to implement  MRT precoding. MRT is a popular precoding scheme for massive MIMO settings, since it reduces the computational complexity greatly as compared to zero-forcing and regularized zero-forcing precoding, which involve the inversion of the Gram matrix of joint users' channel matrix. With MRT, the precoding vector is given as ${\mathbf{g}}_{k} = \zeta \hat {\mathbf{h}}_{k,\mathbf{v}}$, where $\zeta $ satisfies the power constraint in (\ref{p_cons}) as,
\begin{align}
\zeta^2 = P_{max}/\Psi,
\end{align}
where $\Psi =\mathbb{E}[{\rm{tr}\left( {\mathbf{P}\hat {\mathbf{H}}_{\mathbf{v}}}{ \hat {\mathbf{H}}_{\mathbf{v}}^H} \right)}]$ and $\hat{\mathbf{H}}_{\mathbf{v}}^H=[\hat {\mathbf{h}}_{1,\mathbf{v}}, \hat {\mathbf{h}}_{2,\mathbf{v}} \dots \hat {\mathbf{h}}_{K,\mathbf{v}}]\in \mathbb{C}^{M\times K}$. 

Under MRT precoding, the SINR of user $k$ is defined as
\begin{align}
\label{SINR_MRT}
& \gamma_{k}=\frac{p_{k} |\mathbf{h}_{k,\mathbf{v}}^{H} \hat{\mathbf{h}}_{k,\mathbf{v}}|^{2}}{\sum_{l\neq k} p_{l} |\mathbf{h}_{k,\mathbf{v}}^{H} \hat{\mathbf{h}}_{l,\mathbf{v}}|^{2}+ \frac{\Psi}{\rho}},
\end{align}
where $\rho=\frac{P_{max}}{\sigma^2}$. The individual users' rates $R_k$s are given as $R_k=\log_2(1+\gamma_{k})$ and the sum-rate $R_{sum}$ is defined as
\begin{align}
\label{rate_11}
R_{sum}=\sum_{k=1}^{K} R_{k}.
\end{align}

\subsection{Channel Estimation Protocols}
In this section, we propose two channel estimation protocols under the time-division duplexing (TDD) strategy, where the BS exploits channel reciprocity to estimate the downlink channels. The channel coherence period of $\tau$ sec is divided into an uplink training phase of $\tau_{c}$ sec and a downlink transmission phase of $\tau_{d}$ sec. Throughout the training phase, the users transmit mutually orthogonal pilot sequences. After correlating the received training signal with user $k$'s pilot sequence, the BS estimates $\mathbf{h}_{k,\mathbf{v}}$ using the observation vector \cite{ML}
\begin{align}
\label{ch1}
&\mathbf{y}^{tr}_{k,\mathbf{v}}= \mathbf{h}_{k,\mathbf{v}}+ \mathbf{n}_k^{UL}, \hspace{.1in} k=1,\dots, K,
\end{align}
where $\mathbf{n}_k^{UL}\sim \mathcal{CN}(\mathbf{0},\frac{1}{\rho_{tr}} \mathbf{I}_{M})$ is the received noise in the uplink and $\rho_{tr}>0$ is the effective training SNR.

\subsubsection{MMSE-ON/OFF Protocol}

Optimizing IRS parameters using the sum-rate expression in \eqref{rate_11}, where the SINR is given by \eqref{SINR_MRT}, requires estimates of  $\mathbf{h}_{d,k}$  as well as $\mathbf{H}_{1}$ and $\mathbf{h}_{2,k}$. This is very challenging since the IRS has no radio resources to transmit or receive pilot symbols for estimating $\mathbf{H}_{1}$ and $\mathbf{h}_{2,k}$.  Therefore, the BS has to estimate all the channels and share the required IRS configuration with the IRS controller. One way to obtain all this CSI is the LS-ON/OFF protocol proposed in \cite{LS} for a single-user system. We extend this to the multi-user system under MMSE estimation, which is well-known to achieve a lower mean squared error (MSE).

  To this end, note that  $\mathbf{H}_{1}\boldsymbol{\Theta} \mathbf{h}_{2,k}=\mathbf{H}_{0,k}\mathbf{v}$,  where $\mathbf{H}_{0,k}= \mathbf{H}_{1}\text{diag}(\mathbf{h}_{2,k}^T)$. Each vector $\mathbf{h}_{0,i,k} \in \mathbb{C}^{M\times 1}$ in $\mathbf{H}_{0,k}=[\mathbf{h}_{0,1,k}, \dots, \mathbf{h}_{0,N,k}]$ (shown in red arrows in Fig. \ref{LIS_model}) can be interpreted as the channel from user $k$ to the BS through the IRS when only element $i$ of the IRS is ON i.e. $\alpha_{i}=1, \theta_{i}=0$ and $\alpha_{n}=0$, $\forall n\neq i$. Therefore we can represent (\ref{ch1}) as
\begin{align}
\label{train}
&\mathbf{y}_{k,\mathbf{v}}^{tr}=(\mathbf{h}_{d,k}+\sum_{i=1}^{N} \mathbf{h}_{0,i,k}v_{i})+\mathbf{n}_k^{UL}.
\end{align}

To obtain MMSE estimates of $\mathbf{h}_{0,i,k}$ and $\mathbf{h}_{d,k}$, the channel estimation interval is divided into $N+1$ sub-phases of length $ \tau_{s}=\frac{\tau_{c}}{N+1}$ sec. During the first sub-phase, all IRS elements are turned OFF, i.e. $\alpha_n=0$, $\forall n$ and the BS obtains $\hat{\mathbf{h}}_{d,k}$ using the observation vector $\mathbf{y}^{tr}_{1,k}= \mathbf{h}_{d,k}+ \mathbf{n}_{1,k}^{UL}$.  In the following $(i+1)^{th}$ sub-phase, where $i=1,\dots,N$, only the $i^{th}$ IRS element is turned ON in the full reflection mode (i.e. $\alpha_{i}=1$, $\theta_{i}=0$) and BS estimates $\mathbf{h}_{0,i,k}$ using the observation vector $\tilde{\mathbf{y}}^{tr}_{i,k}= \tilde{\mathbf{h}}_{d,k}+  \mathbf{h}_{0,i,k}+\mathbf{n}_{i,k}^{UL}$, obtained after subtracting the contribution of $\hat{\mathbf{h}}_{d,k}$ from $\mathbf{y}^{tr}_{i,k}$, where $\tilde{\mathbf{h}}_{d,k}= \mathbf{h}_{d,k}-\hat{\mathbf{h}}_{d,k}$.  The channels estimates are expressed in the following lemma. 

\begin{lemma} \label{L1} The MMSE estimate of $\mathbf{h}_{k,\mathbf{v}}$ is given as 
\begin{align}
\label{est_corr11}
\hat{\mathbf{h}}_{k,\mathbf{v}}= \hat{\mathbf{h}}_{d,k}+\sum_{i=1}^{N} \hat{\mathbf{h}}_{0,i,k}v_{i},
\end{align}
where the MMSE estimates of $\mathbf{h}_{d,k}$ and $\mathbf{h}_{0,i,k}$ are
\begin{align}
\label{h_d_est}
&\hat{\mathbf{h}}_{d,k}= \mathbf{R}_{d,k}\mathbf{Q}_{d,k}\mathbf{y}^{tr}_{1,k}, \\
\label{h_irs_est}
&\hat{\mathbf{h}}_{0,i,k}= \mathbf{R}_{0,i,k}\mathbf{Q}_{i,k}\tilde{\mathbf{y}}^{tr}_{i,k}, \hspace{.1in} i=1,\dots, N, 
\end{align}
where $ \mathbf{R}_{d,k}=\beta_{d,k}\mathbf{I}_{M}$, $\mathbf{Q}_{d,k}=\frac{\mathbf{I}_{M}}{\beta_{d,k}+1/\rho_{tr}}$, $\mathbf{R}_{0,i,k}=\beta_{2,k}\mathbf{h}_{1,i}\mathbf{h}_{1,i}^{H}$, $\mathbf{Q}_{i,k}=\left(\mathbf{C}_{\tilde{\mathbf{h}}_{d,k}\tilde{\mathbf{h}}_{d,k}^H}+\beta_{2,k}\mathbf{h}_{1,i}\mathbf{h}_{1,i}^{H}+\frac{1 }{\rho_{tr}}\mathbf{I}_M\right)^{-1}$, and $\mathbf{C}_{\tilde{\mathbf{h}}_{d,k}\tilde{\mathbf{h}}_{d,k}^H}=\mathbf{R}_{d,k}-\mathbf{R}_{d,k}\mathbf{Q}_{d,k}\mathbf{R}_{d,k}$.
\end{lemma}
\begin{IEEEproof}
The proof is provided in Appendix \ref{Sec:L1}. 
\end{IEEEproof}

Using these results, we can show that $\hat{\mathbf{h}}_{k,\mathbf{v}}$ behaves as a correlated Rayleigh channel as described below.

\begin{lemma} \label{L3} The channel estimate $\hat{\mathbf{h}}_{k,\mathbf{v}}$ is distributed as
\begin{align}
\label{est_corr}
& \hat{\mathbf{h}}_{k,\mathbf{v}}\sim \mathcal{CN}(\mathbf{0}_M, \mathbf{C}_{k,\mathbf{v}}),
\end{align}
where  \small
\begin{align}
\label{C_k}
&\mathbf{C}_{k,\mathbf{v}}=\mathbf{R}_{d,k}\mathbf{Q}_{d,k}\mathbf{R}_{d,k}+\sum_{i=1}^{N} \alpha_i^2 \mathbf{R}_{0,i,k}\mathbf{Q}_{i,k}\Big( \mathbf{R}_{0,i,k}+\frac{\mathbf{I}_M} {\rho_{tr}}\Big)\nonumber \\
&\mathbf{Q}_{i,k}^H \mathbf{R}_{0,i,k}^H+\sum_{i=1}^{N} \sum_{j=1}^{N} v_{i} \mathbf{R}_{0,i,k}\mathbf{Q}_{i,k}\mathbf{C}_{\tilde{\mathbf{h}}_{d,k}\tilde{\mathbf{h}}_{d,k}^{H}}\mathbf{Q}_{j,k}^H \mathbf{R}_{0,j,k}^H v_{j}^*.
\end{align} \normalsize
 \end{lemma}
 \begin{IEEEproof}
The proof follows from noting that $\hat{\mathbf{h}}_{k,\mathbf{v}}$ is a sum of Gaussian vectors and is omitted due to limited space. 
\end{IEEEproof}

\subsubsection{MMSE-Direct Estimation}

In the DE scheme, instead of estimating the individual channels $\mathbf{h}_{d,k}$ and $\mathbf{h}_{0,i,k}$s, the BS directly estimates   the overall channel $\mathbf{h}_{k,\mathbf{v}}$ for each user for a given $\mathbf{v}$. This is done in a single sub-phase of length $\tau_c=\tau_s$ sec, using the training signal in \eqref{ch1}. The impact of the choice of $\mathbf{v}$ during channel estimation on the sum-rate performance will be studied in Sec. III in the large system limit. The MMSE estimate of $\mathbf{h}_{k,\mathbf{v}}$, for any given $\mathbf{v}$, is now stated.

\begin{lemma} \label{L4} The MMSE estimate $\hat{\mathbf{h}}_{k,\mathbf{v}}$ under DE is
\begin{align}
\label{est_de}
& \hat{\mathbf{h}}^{DE}_{k,\mathbf{v}}= \mathbf{R}_{k,\boldsymbol{\alpha}} \mathbf{Q}_{k,\boldsymbol{\alpha}} \mathbf{y}_{k,\mathbf{v}}^{tr},
\end{align}
where $\mathbf{R}_{k,\boldsymbol{\alpha}}=\beta_{d,k}\mathbf{I}_M +\beta_{2,k} \mathbf{H}_1 \text{diag}(\boldsymbol{\alpha})^{2}\mathbf{H}_1^H$, $\mathbf{Q}_{k,\boldsymbol{\alpha}}=\left(\beta_{d,k}\mathbf{I}_M +\beta_{2,k} \mathbf{H}_1 \text{diag}(\boldsymbol{\alpha})^{2} \mathbf{H}_1^H +\frac{ \mathbf{I}_M}{\rho_{tr}} \right)^{-1}$, $\boldsymbol{\alpha}=[\alpha_1, \dots, \alpha_N]^T$ and $\mathbf{y}_{k,\mathbf{v}}^{tr}$ is given by \eqref{ch1}.
 \end{lemma}
\begin{IEEEproof}
The proof uses the fact that $\boldsymbol{\Theta}\boldsymbol{\Theta}^H=\text{diag}(\boldsymbol{\alpha})^2$, since $|v_n|^2=|\alpha_n \exp(j\theta_n)|^2=\alpha_n^2$. 
\end{IEEEproof}

\begin{remark}  Denoting the estimation error as $\tilde{\mathbf{h}}^{DE}_{k,\mathbf{v}}=\mathbf{h}_{k,\mathbf{v}}-\hat{\mathbf{h}}^{DE}_{k,\mathbf{v}}$, the normalized MSE, $\text{NMSE}(\hat{\mathbf{h}}^{DE}_{k,\mathbf{v}})$, is given as
\begin{align}
\label{DE}
&\frac{\text{tr}(\mathbb{E}[\tilde{\mathbf{h}}^{DE}_{k,\mathbf{v}} \tilde{\mathbf{h}}_{k,\mathbf{v}}^{DE^H}])}{\text{tr}(\mathbb{E}[\mathbf{h}_{k,\mathbf{v}}\mathbf{h}_{k,\mathbf{v}}^H])}=\frac{\text{tr}(\mathbf{R}_{k,\boldsymbol{\alpha}}-\mathbf{R}_{k,\boldsymbol{\alpha}} \mathbf{Q}_{k,\boldsymbol{\alpha}} \mathbf{R}_{k,\boldsymbol{\alpha}}^H)}{\text{tr} \mathbf{R}_{k,\boldsymbol{\alpha}}}.
\end{align}
Interestingly, the NMSE in the estimated channel is independent of the phase shifts $\theta_n$'s and only depends on $\alpha_n$'s.
\end{remark}


The DE scheme reduces the channel estimation time $\tau_c$ from $(N+1)\tau_s$ sec to $\tau_s$ sec, and will therefore significantly reduce the rate loss due to channel training. The downside to this protocol is that although the estimate in \eqref{est_de} can be used to implement the precoding at the BS, it can not be used to design the IRS phases based on the exact sum-rate expression in \eqref{rate_11}, which will require the estimates $\hat{\mathbf{h}}_{d,k}$ and $\hat{\mathbf{h}}_{0,i,k}$. However, if the IRS can be designed using knowledge of only the channel statistics as will be the case in the next section which focuses on large systems, then DE is a very desirable scheme. This is because CSI will only be needed for implementing MRT precoding, for which the BS can use \eqref{est_de} instead of \eqref{est_corr11}.

\section{Asymptotic Analysis} \label{Sec:Asym}

In this section we derive the deterministic approximations of users' SINRs and rates.



\subsection{Main Results}

With the deployment of large-dimension systems in 5G, performance analysis under the asymptotic regime has become an area of interest. In the large $M,N,K$ limit, the users' SINRs and ergodic rates tend to approach deterministic equivalents, which are almost surely (a.s.) tight and depend only on the slowly-varying statistics of the channel. They have been useful in solving important optimization problems in \cite{Wag,  ourworkTCOM, ML2}, motivated by the fact that the optimized solution will only require knowledge of channel statistics instead of instantaneous CSI. In this section, we derive the deterministic approximations of the users' SINR in \eqref{SINR_MRT} and sum-rate in \eqref{rate_11}, which will later yield important insights into the impact of IRS. The analysis requires following  assumptions.

\begin{assumption}\label{A1}
$M$, $N$ and $K$ grow large with a bounded ratio as $0< \liminf_{M,K \rightarrow \infty} \frac{K}{M}\leq \limsup_{M,K \rightarrow \infty}\frac{K}{M}<\infty$ and $0< \liminf_{M,N \rightarrow \infty} \frac{M}{N}\leq \limsup_{M,N \rightarrow \infty}\frac{M}{N}<\infty$. 
\end{assumption}

\begin{assumption}\label{A2}
$\mathbf{H}_1$ satisfies $\limsup_{M} ||\mathbf{H}_1 \mathbf{H}_1^H||<\infty$.
\end{assumption}

\begin{assumption}\label{A3}
The  powers $p_{1}, \dots, p_{K}$ are of order $O(1/K)$.
\end{assumption}

We present the deterministic equivalent of the SINR under MMSE-ON/OFF protocol in the theorem below.

\begin{theorem}\label{Thm1} 
  Under Assumptions \ref{A1}, \ref{A2} and \ref{A3}, the SINR of user $k$ defined in (\ref{SINR_MRT}) under MMSE ON/OFF channel estimate in \eqref{est_corr11} converges as
\begin{align}
\label{det_SINR1}
\gamma_{k}^{\text{ON/OFF}}-{\gamma}_{k}^{\text{ON/OFF}^\circ}\xrightarrow[M,N,K\rightarrow \infty]{a.s.} 0
\end{align}
where ${\gamma}_{k}^{\text{ON/OFF}^\circ}$ is defined in \eqref{det_SINR}, $\mathbf{C}_{k,\mathbf{v}}$ is defined in \eqref{C_k} and $\mathbf{R}_{d,k}$, $\mathbf{Q}_{d,k}$, $\mathbf{R}_{0,i,k}$, $\mathbf{Q}_{i,k}$, $\mathbf{C}_{\tilde{\mathbf{h}}_{d,k}\tilde{\mathbf{h}}_{d,k}}$ are defined in Lemma \ref{L1}.
\begin{figure*}[ht]
\begin{align}
\label{det_SINR}
{\gamma}_{k}^{\text{ON/OFF}^\circ}= \frac{\frac{p_{k}}{K}\left \lvert\sum_{i=1}^{N}{v_i tr(\mathbf{C}_{\tilde{\mathbf{h}}_{d,k}\tilde{\mathbf{h}}_{d,k}}\mathbf{R}_{0,i,k}\mathbf{Q}_{i,k})}+ \frac{\beta_{d,k}^2M}{\beta_{d,k}+\frac{1}{\rho_{tr}}}+\sum_{i=1}^{N}\alpha_i^2 tr(\mathbf{R}_{0,i,k}\mathbf{R}_{0,i,k}\mathbf{Q}_{i,k})\right \rvert^{2}}{\sum_{l\neq k}{\frac{p_{l}}{K}tr(\mathbf{C}_{l,\mathbf{v}}(\beta_{d,k}\mathbf{I}_M +\beta_{2,k} \mathbf{H}_1 \text{diag}(\boldsymbol{\alpha})^2\mathbf{H}_1^H)})+\frac{\sum_{k=1}^{K}\frac{p_{k}}{K}tr(\mathbf{C}_{k,\mathbf{v}})}{\rho}}
\end{align} 
\hrule
\end{figure*}
\end{theorem}
\begin{IEEEproof}
The proof is provided in Appendix B.
\end{IEEEproof}
Two important insights can be drawn from the expression in \eqref{det_SINR}. First, the phase shifts applied by the IRS elements, i.e. $\theta_n$'s, do not appear anywhere except for the terms involving the error covariance matrix $\mathbf{C}_{\tilde{\mathbf{h}}_{d,k}\tilde{\mathbf{h}}_{d,k}}$ of $\tilde{\mathbf{h}}_{d,k}$, which are multiplied by $v_n$'s in \eqref{C_k} and \eqref{det_SINR}. The error in the estimation of $\mathbf{h}_{d,k}$s propagates to the estimation of the subsequent $\mathbf{h}_{0,i,k}$s under the ON/OFF protocol as shown in \eqref{h_irs_est} making $\mathbf{h}_{d,k}$ and $\tilde{\mathbf{y}}^{tr}_{i,k}$, $i=1,\dots , N$ dependent due to the presence of $\tilde{\mathbf{h}}_{d,k}$ in $\tilde{\mathbf{y}}^{tr}_{i,k}$. If the direct channel is estimated accurately, the values of phase shifts will not matter much asymptotically  resulting in a small reflect beamforming gain. This phenomenon is caused by the spatial isotropy that holds upon the IRS-assisted channel, which is insensitive to the beamforming between $\mathbf{H}_1$ and $\mathbf{h}_{2,k}$ under Rayleigh fading $\mathbf{h}_{2,k}$s. Second, the IRS still yields an array gain due to the sum over $N$ in the numerator. We will get insight into the extent of this gain later in this section. 

The deterministic equivalent under DE is now presented. 

\begin{theorem}\label{Thm2} 
  Under Assumptions \ref{A1}, \ref{A2} and \ref{A3}, the SINR of user $k$ defined in (\ref{SINR_MRT}) under DE protocol in \eqref{est_de} converges as $\gamma^{DE}_{k}-{\gamma}_{k}^{\text{DE}^\circ}\xrightarrow[M,N,K\rightarrow \infty]{a.s} 0$, where ${\gamma}_{k}^{DE^\circ}$ is given as ${\gamma}_{k}^{\text{DE}^\circ}=$
\begin{align}
\label{det_SINR_DE}
&\frac{\frac{p_{k}}{K}tr(\mathbf{R}_{k,\boldsymbol{\alpha}} \mathbf{R}_{k,\boldsymbol{\alpha}}\mathbf{Q}_{k,\boldsymbol{\alpha}})^2}{\sum_{l\neq k} \frac{p_{l}}{K}tr(\mathbf{R}_{k,\boldsymbol{\alpha}} \mathbf{R}_{l,\boldsymbol{\alpha}} \mathbf{Q}_{l,\boldsymbol{\alpha}} \mathbf{R}_{l,\boldsymbol{\alpha}})+\sum_{k=1}^{K}\frac{\frac{p_{k}}{K} tr(\mathbf{R}_{k,\boldsymbol{\alpha}} \mathbf{R}_{k,\boldsymbol{\alpha}} \mathbf{Q}_{k,\boldsymbol{\alpha}})}{\rho}},
\end{align} 
where $\mathbf{R}_{k,\boldsymbol{\alpha}}$ and $\mathbf{Q}_{k,\boldsymbol{\alpha}}$ are defined in Lemma \ref{L4}.
\end{theorem}
\begin{IEEEproof}
The proof is similar to the one in Appendix B.
\end{IEEEproof}

Under DE and Rayleigh fading IRS-to-user channels, we see through the expression in \eqref{det_SINR_DE} that the values of IRS  phase-shifts $\theta_n$'s do not affect the SINR asymptotically, which is only affected by $\alpha_n$'s. Therefore  for large systems under the DE protocol, the IRS can adopt any values for the phase-shifts during channel estimation  as well as during downlink transmission without affecting the sum-rate.  However, the IRS will still yield an array gain, which will be explicitly studied later in this section. Next we express the deterministic equivalents of user rates in the following corollary.

\begin{corollary} \label{Cor_rate}
Under Assumptions \ref{A1}, \ref{A2} and \ref{A3}, it follows from the continuous mapping theorem that the individual downlink rates $R_{k}$ of the users converge as $R_{k}-R_{k}^{\circ}\xrightarrow[M,N,K\rightarrow \infty]{a.s} 0$, where $R_{k}^{\circ}= \log(1+{\gamma}_{k}^{\circ})$ and ${\gamma}_{k}^{\circ}$ is given by (\ref{det_SINR}) or \eqref{det_SINR_DE} depending on the estimation protocol. An approximation for the average sum rate can be obtained as
\begin{equation}
\label{R_sum}
R_{sum}^{\circ}= \sum_{k=1}^{K}\log(1+{\gamma}_{k}^{\circ}).
\end{equation}
\end{corollary}
%
%
Finally, we simplify Theorem \ref{Thm1} and \ref{Thm2} for perfect CSI case.

\begin{corollary} \label{Cor_per} Under perfect CSI for both protocols, we have 
\begin{align}
&\gamma_{k}^{\circ}=\frac{\frac{p_{k}}{K}|tr(\mathbf{R}_{k,\boldsymbol{\alpha}})|^{2}}{\sum_{l\neq k}\frac{p_{l}}{K}tr(\mathbf{R}_{l,\boldsymbol{\alpha}}\mathbf{R}_{k,\boldsymbol{\alpha}})+\frac{\frac{p_{k}}{K}\sum_{k=1}^{K}{tr(\mathbf{R}_{k,\boldsymbol{\alpha}})}}{\rho}}.
\end{align}
 \end{corollary}

\vspace{-.02in}
\subsection{Optimization of $\mathbf{v}$ in Theorem \ref{Thm1}}

Under the MMSE-ON/OFF protocol, the phase shifts appear in some terms of the deterministic equivalent in Theorem 1. Denoting $v_n=\alpha_n \tilde{v}_n$, where $\tilde{v}_n=\exp(j\theta_n)$, we formulate the following optimization problem \textit{(P1)}. 
\begin{align} 
&\underset{\tilde{v}_1, \dots, \tilde{v}_N}{\text{max }}  R_{sum}^{\circ}=\sum_{k=1}^{K} \log_2(1+\gamma_{k}^{\text{ON/OFF}^\circ}), \text{s.t. } |\tilde{v}_{n}|=1,  \forall n. \nonumber
\end{align}

\textit{(P1)} is a constrained maximization problem that can be solved using projected gradient ascent. In each step we project the solution to the closest feasible point that satisfies the constraint. The method is explained in Algorithm \ref{alg:euclid} and the derivative of $R_{sum}^{\circ}$ with respect to $\tilde{v}_{n}$, $n=1, \dots N$ is 

\small
\begin{align}
\label{der_rate}
&\frac{\partial R_{sum}^{\circ}}{\partial \tilde{v}_{n}}=\sum_{k=1}^{K}\frac{2 d_k \sqrt{\frac{p_{k}q_{k}}{K}}\alpha_n \text{tr}(\textbf{C}_{\tilde{\textbf{h}}_{d,k}\tilde{\textbf{h}}_{d,k}}\textbf{R}_{0,n,k}\textbf{Q}_{n,k})- q_k d_k'}{(1+\gamma_{k}^o)\ln(2) d_k^2}, \nonumber \\
&d'_k=2\alpha_n\sum_{i=1}^N v_i \Big(\sum_{l\neq k} \frac{p_{l}}{K} tr(\mathbf{R}_{0,i,l} \mathbf{Q}_{i,l} \mathbf{C}_{\tilde{\textbf{h}}_{d,l}\tilde{\textbf{h}}_{d,l}} \mathbf{Q}_{n,l}^H \mathbf{R}_{0,n,l}^H \mathbf{R}_{k,\boldsymbol{\alpha}})\nonumber \\
&+\sum_{k=1}^K\frac{p_k}{K \rho}  tr(\mathbf{R}_{0,i,k} \mathbf{Q}_{i,k} \mathbf{C}_{\tilde{\textbf{h}}_{d,k}\tilde{\textbf{h}}_{d,k}} \mathbf{Q}_{n,k}^H \mathbf{R}_{0,n,k}^H )\Big),
\end{align}\normalsize
where $q_k$ and $d_k$ are numerator and denominator of \eqref{det_SINR} respectively.

\begin{algorithm}[!t]
\caption{Projected Gradient Ascent Algorithm}\label{alg:euclid}
\begin{algorithmic}[1]
\State \textbf{Initialize:}  $s=1$, $\textbf{v}^s$, $R_{sum}^{\circ^s}=f(\mathbf{v}^s)$ in (\ref{R_sum}) and $\epsilon>0$;
\State  \textbf{Repeat}
\State $[\textbf{p}^{s}]_{n}=\frac{\partial R_{sum}^{\circ}}{\partial \tilde{v}_{n}}|_{\mathbf{v}^s}$, $n=1,\dots, N$;
\State \textbf{Find} step size $\mu$ using backtrack line search;
\State $\bar{\textbf{v}}^{s+1}=\tilde{\textbf{v}}^{s}+\mu \textbf{p}^{s}$; $\textbf{v}^{s+1}=\boldsymbol{\alpha}\circ \exp(j \text{arg } (\bar{\textbf{v}}^{s+1}))$;
\State $R_{sum}^{o^{s+1}}=f(\mathbf{v}^{s+1})$ ; 
\State $s=s+1$;
\State \textbf{Until} $||R_{sum}^{o^{s}}-R_{sum}^{o^{s-1}}||^2< \epsilon$; 
\State \textbf{Obtain} $\mathbf{v}^*=\mathbf{v}^{s}$;
\end{algorithmic}
\end{algorithm}
%
%
%

\subsection{How Useful is the IRS?}

To gain explicit insights into the impact of IRS, we consider a special case which assumes $\mathbf{H}_1=\sqrt{\beta_1 N}\textbf{U}$, where $\textbf{U}\in \mathbb{C}^{M\times N}$ is composed of $M\leq N$ leading rows of an arbitrary $N\times N$ unitary matrix \cite{ML}. Since in practice each diagonal entry of $\mathbf{H}_1\mathbf{H}_1^H$ is the sum of $N$ exponential terms of unit norm, so we have normalized $\mathbf{U}$ by  $\sqrt{N}$. Moreover $N$ is assumed to be large but fixed to ensure a bounded spectral norm. The model implies that $\mathbf{H}_1$ has orthogonal rows. Such a LoS scenario can be realized in practice through proper placement of the IRS array with respect to the BS array as discussed in \cite{gesbert}. In any case, this special case will act as an upper-bound on the IRS performance under arbitrary $\mathbf{H}_1$. 

 Moreover, we let $\beta_{1} \beta_{2,k}=c\beta_{d,k}$, $\forall k$, which is justified in scenarios where IRS is located very close to the BS. For this special case, the performance of the IRS-assisted system under perfect CSI is given in a compact closed-form as follows.

\begin{corollary}\label{Cor:spec} For this special case, ${\gamma}_{k}^{\circ}$ in Corollary \ref{Cor_per} under  $p_k=\frac{1}{K}$, $\forall k$ is given as
\begin{align}
\label{special1}
&{\gamma}_{k}^{\circ}=\frac{1}{\underbrace{\frac{1}{M}\sum_{l\neq k} \frac{\beta_{d,l}}{\beta_{d,k}}}_{\text{Interference}}+\underbrace{\frac{\sum_{l=1}^K \beta_{d,l}}{M \beta_{d,k}^2\rho (cN+1)}}_{\text{Noise}}}.
\end{align}
\end{corollary}

This corollary yields two important insights. First it verifies the ``massive MIMO effect" observed in \cite{ML}, that as $M$ increases for fixed $N$ and $K$, the user rates grows logarithmically. Second, the use of an IRS under Rayleigh fading $\textbf{h}_{2,k}$s is only useful in large systems when the average received SNR is low, i.e. either $\rho$ is low or the path loss is high which results in small $\beta_{d,k}$. This is often the case for cell-edge users. In such a noise-limited scenario\footnote{Noise-limited here implies that the noise power is much higher than the intra-cell interference power. In multi-cell systems (not the focus of this work), the total noise consists of the noise plus inter-cell interference.}, the second term in the denominator of \eqref{special1} will dominate the first and increasing $N$ will produce a noticeable increase in the SINR values. In an interference-limited scenario, the use of an IRS yields no substantial benefit. This can be intuitively explained by noting that under Rayleigh fading IRS-to-user channels, the IRS only yields an array gain of $N$ asymptotically. This gain appears in both the energy of desired and interfering signals and the net effect becomes negligible if the the interference is dominant. 


It will be interesting to extend the asymptotic analysis to Rician fading in the future, where the IRS phase-shifts will play a significant role, resulting in a reflect beamforming gain. 

\vspace{-.05in}
\section{Simulation Results}\label{Sec:Sim}
\vspace{-.03in}

We consider $K$ cell-edge users placed  uniformly (with constant angular gap) along an arc of radius $150$m that spans angles from $-30^o$ to $30^o$ with respect to the $x$-axis. Using $(x,y,z)$ coordinates (in meters), the BS and IRS are deployed at $(0,0,25)$ and $(\bar{x},0,40)$ respectively where $\bar{x}$ is the mean of $x$ coordinates of all users. We assume a full rank BS-to-IRS LoS channel $\mathbf{H}_1$ from \cite{annie}, and define  $p_k=1/K$, $\forall k$. The considered path loss  model is $\beta=\frac{C_0}{d^{\eta}}$, where $C_0=30$\rm{dB} and the path loss exponent $\eta$ is  set as $2$ for $\mathbf{H}_1$, $2.8$ for $\mathbf{h}_{2,k}$ and $3.5$ for $\mathbf{h}_{d,k}$ \cite{Wu1, Wu_dis}.  We set  $\alpha_{n}=1$  $\forall n$, similar to almost all existing works on IRS-assisted systems \cite{Guo, Wu1, annie, huang, LS}.



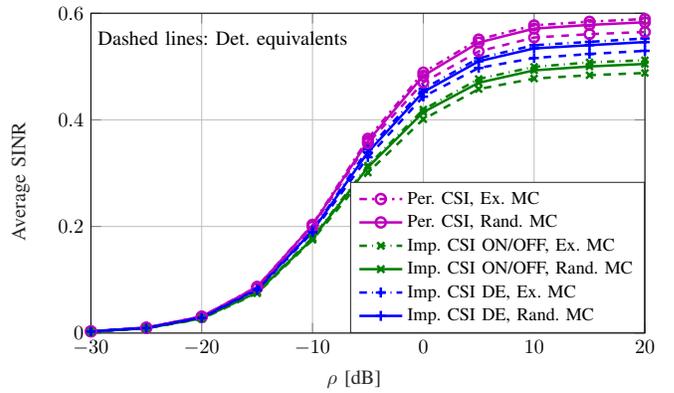
\begin{figure}[!t]
\tikzset{every picture/.style={scale=.95}, every node/.style={scale=.8}}
%
%
\definecolor{mycolor1}{rgb}{0.74902,0.00000,0.74902}%
\definecolor{mycolor2}{rgb}{0.00000,0.49804,0.00000}%
\begin{tikzpicture}

\begin{axis}[%
width=.45\textwidth,
height=.26\textwidth,,
scale only axis,
xmin=-30,
xmax=20,
xlabel style={font=\color{white!15!black}},
xlabel={$\rho\text{ [dB]}$},
ymin=0,
ymax=0.6,
ylabel style={font=\color{white!15!black}},
ylabel={Average SINR},
axis background/.style={fill=white},
xmajorgrids,
ymajorgrids,
legend style={at={(axis cs: 20,0)},anchor=south east,legend cell align=left,align=left,draw=white!15!black, /tikz/column 2/.style={
                column sep=5pt,
            }}
						]
\addplot [color=mycolor1, line width=1.0pt, mark size=2.0pt, dashdotted, mark=o, mark options={solid, mycolor1}]
  table[row sep=crcr]{%
-30	0.00328493842849\\
-25	0.01020485783929\\
-20	0.03104877372848\\
-15	0.086949204857\\
-10	0.20347839572386\\
-5	0.3647493957394\\
0	0.48894728572847\\
5	0.550958365937495\\
10	0.57735837598294\\
15	0.5842948568356\\
20	0.589194869759385\\
};
\addlegendentry{\small Per. CSI, Ex. MC}

\addplot [color=mycolor1, line width=1.0pt, mark size=2.0pt, mark=o, mark options={solid, mycolor1}]
  table[row sep=crcr]{%
-30	0.00327983471590012\\
-25	0.0101877559745347\\
-20	0.0309630573371485\\
-15	0.0863762439032278\\
-10	0.202369444159586\\
-5	0.359884381107703\\
0	0.48270114361151\\
5	0.545105455449041\\
10	0.571147417463231\\
15	0.578162922997759\\
20	0.583083558785541\\
};
\addlegendentry{\small Per. CSI, Rand. MC}

\addplot [color=mycolor2, dashdotted, line width=1.0pt, mark size=2.0pt,mark=x, mark options={solid, mycolor2}]
  table[row sep=crcr]{%
-30	0.003000753628474\\
-25	0.00927485732749\\
-20	0.02813852747382\\
-15	0.0779483734567\\
-10	0.178978573936\\
-5	0.31416857337583\\
0	0.4193756274628585\\
5	0.47667483958394\\
10	0.49945834895829\\
15	0.50784939573559\\
20	0.5118439583938595\\
};
\addlegendentry{\small Imp. CSI ON/OFF, Ex. MC}

\addplot [color=mycolor2, line width=1.0pt, mark size=2.0pt, mark=x, mark options={solid, mycolor2}]
  table[row sep=crcr]{%
-30	0.0029387598890298\\
-25	0.00910797401248723\\
-20	0.0276116138917449\\
-15	0.0764193309239968\\
-10	0.176888689535276\\
-5	0.310904853911627\\
0	0.414437689577715\\
5	0.469615257105466\\
10	0.492909257951506\\
15	0.500202142127164\\
20	0.504861076963611\\
};
\addlegendentry{\small Imp. CSI ON/OFF, Rand. MC}

\addplot [color=blue, dashdotted, line width=1.0pt, mark size=2.0pt,mark=+, mark options={solid, blue}]
  table[row sep=crcr]{%
-30	0.0031575839734\\
-25	0.009889473857394\\
-20	0.029894628739585\\
-15	0.082798375937593\\
-10	0.192895739583\\
-5	0.34472573957392\\
0	0.4578738573853\\
5	0.51567383959294\\
10	0.54003842849382\\
15	0.54629759375937\\
20	0.552738758474\\
};
\addlegendentry{\small Imp. CSI DE, Ex. MC}

\addplot [color=blue, line width=1.0pt, mark size=2.0pt, mark=+, mark options={solid, blue}]
  table[row sep=crcr]{%
-30	0.00313910514059484\\
-25	0.00974514092548416\\
-20	0.0295915073632481\\
-15	0.0823407231636365\\
-10	0.1917784956388\\
-5	0.339034706565292\\
0	0.45230889092929\\
5	0.509826225371026\\
10	0.534061436381705\\
15	0.540094377035284\\
20	0.545956616796545\\
};
\addlegendentry{\small Imp. CSI DE, Rand. MC}

\addplot [color=mycolor1, dashed, line width=1.0pt, mark size=2.0pt,mark=o, mark options={solid, mycolor1}]
  table[row sep=crcr]{%
-30	0.00309571923338553\\
-25	0.00967026008813369\\
-20	0.029945643542573\\
-15	0.084338933758793\\
-10	0.199851404706866\\
-5	0.355987918366856\\
0	0.470184115359238\\
5	0.528397195746914\\
10	0.554454509825665\\
15	0.560778315749238\\
20	0.564809398774249\\
};

\addplot [color=mycolor2, dashed, line width=1.0pt, mark size=2.0pt, mark=x, mark options={solid, mycolor2}]
  table[row sep=crcr]{%
-30	0.00274021938341259\\
-25	0.008853490289385\\
-20	0.026887798433273\\
-15	0.074613834825621\\
-10	0.174905226950926\\
-5	0.301160497845975\\
0	0.401168518070692\\
5	0.457745219235844\\
10	0.477506385835041\\
15	0.483790789196493\\
20	0.487939027413395\\
};

\addplot [color=blue, dashed, line width=1.0pt, mark size=2.0pt, mark=+, mark options={solid, blue}]
  table[row sep=crcr]{%
-30	0.00295121695538312\\
-25	0.009316407544194\\
-20	0.028541140354429\\
-15	0.080919051131761\\
-10	0.18903451482732\\
-5	0.329910751594833\\
0	0.443537952981622\\
5	0.497400909331757\\
10	0.516020921081475\\
15	0.523595582621401\\
20	0.52949674758671\\
};

\node at (axis cs: -30,0.55) [anchor = west] {\normalsize Dashed lines: Det. equivalents };

\end{axis}
\end{tikzpicture}%
\caption{Validation of the deterministic (det.) equivalent of the SINR against the Monte-Carlo (MC) values, under perfect (per.) and imperfect (imp.) CSI.}
\label{equiv_model}
\end{figure}

First, we plot against $\rho$ in Fig. \ref{equiv_model}: 1) the Monte-Carlo simulated average SINR in \eqref{SINR_MRT} under exhaustive (Ex.) search over combinations of phase-shifts implementable using $2$-bit IRS phase shifters \cite{Wu_dis} that maximize the average SINR, 2) the Monte-Carlo simulated average SINR under random (Rand.) phase-shifts from the discrete set, and 3) the deterministic equivalent of average SINR. All three quantities are plotted for perfect CSI, as well as imperfect CSI under MMSE-ON/OFF protocol and  DE protocol, for $M=K=N=48$ and $\rho_{tr}=8$\rm{dB}. The deterministic equivalents under DE scheme and perfect CSI  do not depend on phase-shift values as seen in Theorem 2 and Corollary 2. The phase-shifts in the deterministic equivalent under MMSE-ON/OFF protocol in \eqref{det_SINR} are computed using \textbf{Algorithm 1}. The deterministic equivalents provide a very good approximation to the Monte-Carlo simulated average SINR for moderate system size. Moreover, we see that optimizing IRS phase shifts (i.e. exhaustive search) under Rayleigh fading $\mathbf{h}_{2,k}$s provides a very small performance gain for large systems. This is in accordance with our observation in Sec. III that asymptotically the values of phase-shifts do not matter much and the IRS only yields an array gain.  The DE scheme performs better than the ON/OFF protocol since the estimation of the overall $\mathbf{h}_{k,\mathbf{v}}$ in the latter protocol suffers from errors in the estimation of $N+1$ channel vectors. DE is therefore desirable, provided CSI of individual links is not needed to design the IRS, which is the case for large systems.



Next we introduce the model for $\rho_{tr}$ as $\rho_{tr}=\frac{p_c \tau_c}{\sigma^2}$, which depends on the noise variance, $\sigma^2$, set as $10^{-19}$ Joules, the pilot Tx power $p_c$ set as $1$W and the training period defined as $\tau_c=0.01 \tau$, where $\tau=20$ms. The net achievable rate of user $k$  is $R_k=\left(1-\frac{\tau_{c}}{\tau} \right)\text{log}_{2}(1+\gamma_{k})$ and $R_{sum}=\sum_{k=1}^K R_k$. 



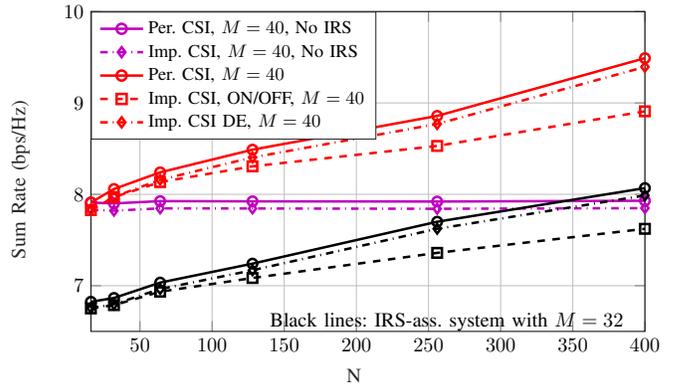
\begin{figure}[!t]
\tikzset{every picture/.style={scale=.95}, every node/.style={scale=.8}}
%
%
\definecolor{mycolor1}{rgb}{0.74902,0.00000,0.74902}%
\begin{tikzpicture}

\begin{axis}[%
width=.45\textwidth,
height=.26\textwidth,,
scale only axis,
xmin=16,
xmax=400,
xlabel style={font=\color{white!15!black}},
xlabel={N},
ymin=6.5,
ymax=10,
ylabel style={font=\color{white!15!black}},
ylabel={Sum Rate (bps/Hz)},
axis background/.style={fill=white},
xmajorgrids,
ymajorgrids,
legend style={at={(axis cs: 16,10)},anchor=north west,legend cell align=left,align=left,draw=white!15!black, /tikz/column 2/.style={
                column sep=5pt,
            }}
						]
\addplot [color=mycolor1,line width=1.0pt, mark size=2.0pt,  mark=o, mark options={solid, mycolor1}]
  table[row sep=crcr]{%
16	7.90998502745228\\
32	7.90040072958807\\
64	7.92560321458496\\
128	7.92334188375321\\
256	7.9210601184157\\
400	7.92970295332452\\
};
\addlegendentry{\small Per. CSI, $M=40$, No IRS}

\addplot [color=mycolor1, dashdotted, line width=1.0pt, mark size=2.0pt, mark=diamond, mark options={solid, mycolor1}]
  table[row sep=crcr]{%
16	7.83109135195497\\
32	7.82116833710003\\
64	7.84640454879196\\
128	7.8449361186599\\
256	7.84184566043095\\
400	7.84883297121641\\
};
\addlegendentry{\small Imp. CSI, $M=40$, No IRS}

\addplot [color=red, mark=o, line width=1.0pt, mark size=2.0pt, mark options={solid, red}]
  table[row sep=crcr]{%
16	7.91037981305129\\
32	8.05829006313738\\
64	8.24028969355427\\
128	8.48861828071053\\
256	8.85803914451762\\
400	9.48985300864461\\
};
\addlegendentry{\small Per. CSI, $M=40$}

\addplot [color=red, dashed, line width=1.0pt, mark size=2.0pt, mark=square, mark options={solid, red}]
  table[row sep=crcr]{%
16	7.8271537191299\\
32	7.96934729496364\\
64	8.13398834672543\\
128	8.30656810169183\\
256	8.52957506761572\\
400	8.908392980892102\\
};
\addlegendentry{\small Imp. CSI, ON/OFF, $M=40$}

\addplot [color=red, dashdotted, line width=1.0pt, mark size=2.0pt, mark=diamond, mark options={solid, red}]
  table[row sep=crcr]{%
16	7.83125980787795\\
32	7.97272087611056\\
64	8.15888985095881\\
128	8.40372525020449\\
256	8.76944342937589\\
400	9.39495308211358\\
};
\addlegendentry{\small Imp. CSI DE, $M=40$}

\addplot [color=black, line width=1.0pt, mark size=2.0pt, mark=o, mark options={solid, black}]
  table[row sep=crcr]{%
16	6.82412413288078\\
32	6.86582623880298\\
64	7.03397420573083\\
128	7.24147617347409\\
256	7.70002359801546\\
400	8.06826739343325\\
};

\addplot [color=black, dashed, line width=1.0pt, mark size=2.0pt, mark=square, mark options={solid, black}]
  table[row sep=crcr]{%
16	6.75154389497057\\
32	6.78741586037778\\
64	6.93473310048221\\
128	7.08438167604328\\
256	7.35943615988184\\
400	7.62261067685568\\
};

\addplot [color=black, dashdotted, line width=1.0pt, mark size=2.0pt, mark=diamond, mark options={solid, black}]
  table[row sep=crcr]{%
16	6.75587017932034\\
32	6.79718046560625\\
64	6.96365177346828\\
128	7.16905079566333\\
256	7.62300831676642\\
400	7.9875887163328\\
};

\node at (axis cs: 390,6.6) [anchor = east] {\normalsize Black lines: IRS-ass. system with $M=32$ };

\end{axis}

\end{tikzpicture}%
\caption{Sum-rate comparison of IRS-assisted (IRS-ass.) and conventional MISO (No IRS) systems.}
\label{IRS_effec1}
\end{figure}

%
In Fig. \ref{IRS_effec1}, we study the sum-rate achieved by $8$ users for varying number of BS antennas and IRS  elements and compare the performance with that of a  conventional MISO system with  $40$ BS antennas. The IRS phase shifts are drawn randomly, given our observation from Fig. 2 that even for moderate system sizes optimizing phase-shifts (using exhaustive search) provides a negligible gain. Although the IRS does not yield any notable reflect beamforming gain asymptotically under Rayleigh fading $\mathbf{h}_{2,k}$s, it does yield a significant array gain because of which the IRS-assisted system with $32$ BS antennas and around $320$ IRS reflecting elements can achieve the same performance as the conventional large MISO system, making it an energy-efficient alternative. The gap between perfect and imperfect CSI curves increases significantly with $N$ under MMSE-ON/OFF protocol as the estimation time increases linearly with $N$, while it stays constant under DE. 

\begin{figure}[!t]
\tikzset{every picture/.style={scale=.95}, every node/.style={scale=.8}}
%
%
\definecolor{mycolor1}{rgb}{0.00000,0.49804,0.00000}%
\begin{tikzpicture}

\begin{axis}[%
width=.45\textwidth,
height=.2\textwidth,
scale only axis,
xmin=-20,
xmax=20,
xlabel style={font=\normalsize\color{white!15!black}},
xlabel={$\rho\text{ [dB]}$},
ymin=-20,
ymax=10,
ylabel style={font=\normalsize\color{white!15!black}},
ylabel={Average SINR [dB]},
axis background/.style={fill=white},
xmajorgrids,
ymajorgrids,
legend style={at={(axis cs: 20,-20)},anchor=south east,legend cell align=left,align=left,draw=white!15!black, /tikz/column 2/.style={
                column sep=5pt,
            }}
						]
						
\addplot [color=mycolor1, line width=1.0pt, mark size=2.0pt, mark=square, mark options={solid, mycolor1}]
  table[row sep=crcr]{%
-20	-15.6549109952379\\
-15	-10.742692372494\\
-10	-6.02583076481947\\
-5	-1.74665506335315\\
0	1.54557778085494\\
5	3.61521145044639\\
10	4.53893837800687\\
15	4.86094758419146\\
20	5.02409329663847\\
};
\addlegendentry{Conv. system, MC}

\addplot [color=mycolor1, dashed, line width=1.0pt, mark size=2.5pt, mark=triangle, mark options={solid, mycolor1}]
  table[row sep=crcr]{%
-20	-15.7799416482315\\
-15	-10.8644138585678\\
-10	-6.12121317335858\\
-5	-1.84579045436739\\
0	1.43422142302313\\
5	3.35042840806978\\
10	4.18790147645158\\
15	4.49027816423386\\
20	4.59045191073867\\
};
\addlegendentry{Corollary 2 from \cite{ML}}

\addplot [color=red, line width=1.0pt, mark size=2.0pt, mark=o, mark options={solid, red}]
  table[row sep=crcr]{%
-20	-1.61209900542249\\
-15	1.65430562191976\\
-10	3.66549196434489\\
-5	4.5357252052572\\
0	4.87366616629324\\
5	5.05157518747261\\
10	5.02575501423955\\
15	5.06648107302837\\
20	5.08244154367215\\
};
\addlegendentry{IRS-ass. system, MC, $N=32$}

\addplot [color=red, dashed, line width=1.0pt, mark size=2.8pt, mark=x, mark options={solid, red}]
  table[row sep=crcr]{%
-20	-1.70295059821393\\
-15	1.52983066300975\\
-10	3.39716787446281\\
-5	4.20576590360383\\
0	4.49632650470075\\
5	4.59240827192138\\
10	4.62323976738892\\
15	4.63303527080416\\
20	4.63613748463001\\
};
\addlegendentry{Corollary \ref{Cor:spec}, $N=32$}

\addplot [color=blue, line width=1.0pt, mark size=2.0pt, mark=diamond, mark options={solid, blue}]
  table[row sep=crcr]{%
-20	0.458794017931727\\
-15	2.9865720407793\\
-10	4.27781087496087\\
-5	4.76986783965534\\
0	4.95370462621038\\
5	5.02443301112687\\
10	5.07287875493873\\
15	4.99807806823951\\
20	5.02730350914199\\
};
\addlegendentry{IRS-ass. system, MC, $N=64$}

\addplot [color=blue, dashed, line width=1.0pt, mark size=2.8pt, mark=asterisk, mark options={solid, blue}]
  table[row sep=crcr]{%
-20	0.3589455665812\\
-15	2.78858156045264\\
-10	3.96376859479159\\
-5	4.41298736259552\\
0	4.56528924103724\\
5	4.61458448709943\\
10	4.63029018286877\\
15	4.63526860363242\\
20	4.63684410693146\\
};
\addlegendentry{Corollary \ref{Cor:spec}, $N=64$}

\end{axis}
\end{tikzpicture}%
\caption{Performance of IRS-assisted system under the special case in Sec. III-C for $M=32$ and $K=12$.}
\label{IRS_effec}
\end{figure}
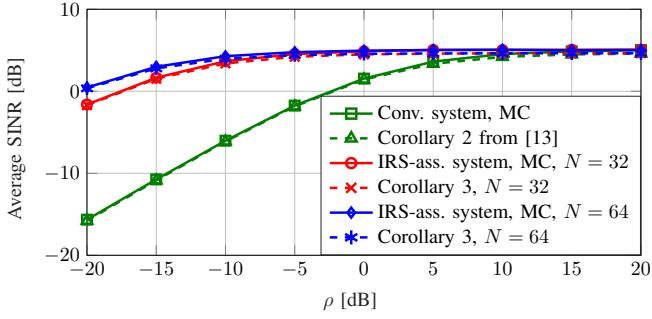

Fig. \ref{IRS_effec} studies the performance of the IRS-assisted system against $\rho=\frac{\beta_{d,k}}{\sigma^2}$ and $c=1$ under the special case in Sec. III-C.  The result in Corollary \ref{Cor:spec} is plotted showing an excellent match between the Monte-Carlo simulated average SINR and the theoretical expression in \eqref{special1}. The IRS is shown to be beneficial under  Rayleigh fading $\mathbf{h}_{2,k}$s, when $\rho$ takes small to moderate values, which is often the case for cell edge or blocked users.  For interference-limited scenarios (i.e. high $\rho$), the performance of IRS-assisted system approaches that of the conventional MISO system under Rayleigh fading, where the latter was studied under this special case in \cite{ML}.

\section{Conclusion}
\label{Sec:Con}

This paper studied the asymptotic performance of an IRS-assisted system under imperfect CSI. We outlined the MMSE-ON/OFF channel estimation protocol which estimates all IRS-assisted links as well as proposed a low-complexity DE scheme, which reduces the channel training time by estimating the overall channel. The deterministic equivalents of SINR and sum-rate were derived under both protocols, which revealed that values of IRS phase-shifts play a negligible role in the large system limit under Rayleigh fading IRS-to-users channels. The theoretical performance  was further studied under a special case to explicitly show that IRS still yields an array gain factor, which becomes very notable in noise-limited systems. Simulation results validated the derived results and highlighted the benefit of relying on DE for precoder design while designing the IRS using statistical information only.

\appendix

\subsection{Proof of Lemma \ref{L1}} \label{Sec:L1}

Given the observed training signal, $\mathbf{y}_{1,k}^{tr}$, we can write the MMSE estimate as $\hat{\mathbf{h}}_{d,k}=\mathbf{W}\mathbf{y}_{1,k}^{tr}$, where $\mathbf{W}$ is the solution to $ \text{min}_{\mathbf{W}}\mathbb{E}[|\hat{\mathbf{h}}_{d,k}-\mathbf{h}_{d,k}|^2]$, which turns out to be  $\mathbf{W}=\mathbb{E}[\mathbf{y}_{1,k}^{tr}\mathbf{h}_{d,k}^H](\mathbb{E}[\mathbf{y}_{1,k}^{tr}\mathbf{y}_{1,k}^{trH}])^{-1}$. Noting that $\mathbf{n}^{UL}_{1,k}$ and $\mathbf{h}_{d,k}$ are independent random vectors we obtain, 
\begin{align}
&\mathbb{E}[\mathbf{y}_{1,k}^{tr}\mathbf{h}_{d,k}^H]=\mathbb{E}[(\mathbf{h}_{d,k}+\mathbf{n}^{UL}_{1,k})\mathbf{h}_{d,k}^H]=\mathbb{E}[\mathbf{h}_{d,k}\mathbf{h}_{d,k}^H]=\beta_{d,k}\mathbf{I}_{M}, \nonumber \\
&\mathbb{E}[\mathbf{y}_{1,k}^{tr}\mathbf{y}_{1,k}^{trH}]=\mathbb{E}[\mathbf{h}_{d,k}\mathbf{h}_{d,k}]+\mathbb{E}[\mathbf{n}^{UL}_{1,k}\mathbf{n}^{UL^H}_{1,k}]=\beta_{d,k}\mathbf{I}_{M}+ \frac{\mathbf{I}_M}{\rho_{tr}} . \nonumber
\end{align}

The proof of \eqref{h_irs_est} follows along similar lines.

\subsection{Sketch of Proof of Theorem 1} \label{Sec:AppB}

We start by dividing the numerator and denominator of (\ref{SINR_MRT}) by $\frac{1}{K}$, resulting in three terms: (i) $\frac{p_k}{K}|\mathbf{h}_{k,\mathbf{v}}^H\hat{\mathbf{h}}_{k,\mathbf{v}}|^2$, (ii) $\sum_{l\neq k}\frac{p_l}{K}|\mathbf{h}_{k,\mathbf{v}}^H\hat{\mathbf{h}}_{l,\mathbf{v}}|^2$,  (iii)  $\frac{1}{K}\Psi$. We show the derivation for (ii). 

Write $\sum_{l\neq k}\frac{p_{l}}{K}|\mathbf{h}_{k,\mathbf{v}}^H\hat{\mathbf{h}}_{l,\mathbf{v}}|^2=\frac{1}{K}\mathbf{h}_{k,\mathbf{v}}^{H}\hat{\mathbf{H}}^{H}_{[k],\mathbf{v}}\mathbf{P}_{[k]}\hat{\mathbf{H}}_{[k],\mathbf{v}}\mathbf{h}_{k,\mathbf{v}}$, where $\hat{\mathbf{H}}^{H}_{[k],\mathbf{v}}$ is $\hat{\mathbf{H}}^{H}_{\mathbf{v}}$ with $k^{th}$ column removed. Note that $\mathbb{E}[\mathbf{h}_{k,\mathbf{v}} \mathbf{h}_{k,\mathbf{v}}^H]=\beta_{d,k} \mathbf{I}_M +\beta_{2,k} \mathbf{H}_1 \boldsymbol{\Theta} \boldsymbol{\Theta}^H \mathbf{H}_1=\beta_{d,k}\mathbf{I}_M +\beta_{2,k} \mathbf{H}_1 \text{diag}(\boldsymbol{\alpha})^2\mathbf{H}_1^H$. We apply trace lemma \cite[Lemma 4]{Wag} to obtain $\frac{1}{K}\mathbf{h}_{k,\mathbf{v}}^{H}\hat{\mathbf{H}}^{H}_{[k],\mathbf{v}}\mathbf{P}_{[k]}\hat{\mathbf{H}}_{[k],\mathbf{v}}\mathbf{h}_{k,\mathbf{v}}-\frac{1}{K}tr((\beta_{d,k}\mathbf{I}_M +\beta_{2,k} \mathbf{H}_1 \text{diag}(\boldsymbol{\alpha})^2 \mathbf{H}_1^H)\hat{\mathbf{H}}^{H}_{[k],\mathbf{v}}\mathbf{P}_{[k]}\hat{\mathbf{H}}_{[k],\mathbf{v}})\xrightarrow[M,K \rightarrow \infty]{a.s.} 0$.

Writing the result as $\sum_{l\neq k}p_{l}\hat{\mathbf{h}}_{l,\mathbf{v}}^{H}(\beta_{d,k}\mathbf{I}_M +\beta_{2,k} \mathbf{H}_1 \text{diag}(\boldsymbol{\alpha})^2 \mathbf{H}_1^H)\hat{\mathbf{h}}_{l,\mathbf{v}}$ and applying trace lemma yields $\frac{1}{K}\sum_{l\neq k}p_{l}\hat{\mathbf{h}}_{l,\mathbf{v}}^{H}(\beta_{d,k}\mathbf{I}_M +\beta_{2,k} \mathbf{H}_1 \text{diag}(\boldsymbol{\alpha})^2 \mathbf{H}_1^H)\hat{\mathbf{h}}_{l,\mathbf{v}}-\frac{1}{K}\sum_{l\neq k}p_{l}tr(\mathbf{C}_{l,\mathbf{v}}(\beta_{d,k}\mathbf{I}_M +\beta_{2,k} \mathbf{H}_1 \text{diag}(\boldsymbol{\alpha})^2 \mathbf{H}_1^H))\xrightarrow[M,K\rightarrow \infty]{a.s} 0$. 


\vspace{-.05in}
\bibliographystyle{IEEEtran}
\bibliography{bibl}

\end{document}